\documentclass[ twocolumn,aps,prd,   
               preprintnumbers,numbers,sort&compress,
               nofootinbib,
                            showpacs,
               colorlinks,
               linkcolor=blue,
               citecolor=blue]{revtex4-1}
   \newcommand{\exclude}[1]{}

\usepackage{graphicx,amsmath,amssymb,bm}
\usepackage{psfrag}
 \usepackage{feynmp}
\usepackage{hyperref}
\usepackage{enumitem}

\newcommand{\beq}{\begin{equation}}
\newcommand{\eeq}{\end{equation}}
\newcommand{\be}{\begin{eqnarray}}
\newcommand{\ee}{\end{eqnarray}}
   
\def\dd{ \,\mathrm{d} }

\def\+{\dagger}
 \def\la{\langle}
 \def\ra{\rangle}

\begin{document}

\title{ Topological  order  and   Berry connection     for the Maxwell  Vacuum on a four-torus  }

\author{   Ariel R. Zhitnitsky}  

\affiliation{ Department of Physics \& Astronomy, University of British Columbia, Vancouver, B.C. V6T 1Z1, Canada} 

\begin{abstract}
We study  novel type of contributions to  the  partition function  of  the Maxwell system  defined on a small compact manifold such as torus. These new terms   can not be described in terms of  the physical propagating photons with two transverse polarizations.  
Rather, these novel contributions  emerge as a result of   tunnelling events  when transitions occur between topologically different but physically  identical vacuum winding states. These new terms give an extra contribution to the Casimir pressure.  The infrared physics in the system can be  described in terms of the topological auxiliary non-propagating fields $a_i(\mathbf{k})$ governed by Chern-Simons -like action. The   system can be studied  in terms of these auxiliary fields precisely in the same way as a topological insulator can be  analyzed  in terms of Berry's connection ${\cal{A}}_i(\mathbf{k})$.  We also  argue that   the Maxwell vacuum defined on a small 4-torus behaves very much in the same way as a topological insulator  with $\theta\neq 0$.

 
\end{abstract} 

\maketitle

\section{Introduction. Motivation.}\label{introduction}
 The main motivation for present studies is as follows.
 It has been recently argued \cite{Cao:2013na,Zhitnitsky:2013hba} that if  the free Maxwell theory (without any interactions with charged particles) is   defined on a small compact manifold  than  some novel  terms in the partition function will emerge.  These terms are not related to the  propagating photons with two transverse physical polarizations, which are responsible for the conventional Casimir effect. Rather, these novel terms occur as a result  of  tunnelling  events between topologically different but physically identical    states. These states play no role when the system is defined in Minkowski space-time ${\mathbb{R}_{1,3}}$. But these states become important when   the system   is defined on a finite compact manifold such as  torus ${\mathbb{T}}^4$. 
 
 In particular,  it has been explicitly shown in  \cite{Cao:2013na,Zhitnitsky:2013hba} that these novel terms lead to a fundamentally new  contributions to the Casimir vacuum pressure, which can  not be expressed  in terms of conventional  propagating  physical degrees of freedom.     Instead,  the new vacuum contributions   appear  as a result  of tunnelling events between different topological sectors $|k\ra$. 
 Mathematically, these sectors  emerge as a result of non-triviality of    the fundamental group $\pi_1[U(1)]\cong \mathbb{Z}$ 
 when the system is defined on a torus.
  
 The crucial for the present studies observation is as follows. While the Maxwell Electrodynamics is the theory of massless particles (photons), the topological portion of the system decouples from dynamics of these massless propagating photons.  Indeed, as we discuss below, the total partition function ${\cal{Z}}$ can be represented as  a product  ${\cal{Z}} ={\cal{Z}}_{0}\times {\cal{Z}}_{\rm top}$.  The conventional partition function ${\cal{Z}}_{0}$ describing physical photons is  not sensitive  to the topological sectors $|k\ra$ of the system which itself is described by ${\cal{Z}}_{\rm top}$. The topological portion of the partition function ${\cal{Z}}_{\rm top}$
 behaves very much   as topological quantum field theory (TQFT) as  argued in \cite{Zhitnitsky:2013hba}. Furthermore, it  demonstrates  many features 
 of   topologically ordered systems, which were  initially   introduced in context of condensed matter (CM) systems, see  original papers \cite{volovik-berry, Wen:1989iv,Wen:1990zza,Moore:1991ks,BF} and recent reviews \cite{Cho:2010rk,Wen:2012hm,Sachdev:2012dq, Cortijo:2011aa, Volovik:2011kg}. 
 
 In particular, ${\cal{Z}}_{\rm top}$ demonstrates  the degeneracy of the system which can not be described in terms of any local operators. Instead, such a degeneracy can be formulated in terms of  some non-local operators  \cite{Zhitnitsky:2013hba}. Furthermore, our system  exhibits some  universal sub-leading corrections  to the thermodynamical entropy which can not be expressed  in terms of  propagating photons with two physical polarizations. Instead,  the corresponding universal contribution to the entropy is expressed in terms of the ``instantons"   describing  the tunnelling events between topologically different but physically identical topological sectors $|k\ra$. 
 
 As a result of these similarities, the key question addressed in the present work is as follows. It has been known for sometime  \cite{Wen:1989iv,Wen:1990zza,Moore:1991ks,BF,Cho:2010rk,Wen:2012hm,Sachdev:2012dq, Cortijo:2011aa, Volovik:2011kg} that some key features of  topologically ordered systems can be formulated in terms of the so-called Berry's connection  in momentum space. Does a   similar description  exist for the Maxwell vacuum defined on a compact manifold?
 
 To address this question we  
 formulate the topological features of the system in terms of an auxiliary  fields. Such a formulation exhibits a close mathematical similarity between the auxiliary topological field describing the Maxwell vacuum state and  
 the  Berry's connection (which is emergent, not a fundamental field)  in topologically ordered CM systems. Such a similarity  looks very instructive and suggestive, and 
   further supports our arguments  \cite{Zhitnitsky:2013hba} that the ground state of the Maxwell theory   defined on a small compact manifold behaves as a TQFT.

  The structure of our presentation is as follows. In the next section \ref{topology}, we review the relevant parts of the two dimensional Maxwell ``empty" theory which does not have  any physical propagating degrees of freedom. Still,  it demonstrates a number of  very  nontrivial topological features present in the system.    In section \ref{4d} we generalize our description  for 
 4d Maxwell theory defined  on four torus.  
 In our main section \ref{berry-section} we introduce the auxiliary fields which effectively account for the  topological sectors  of the system. We study the  behaviour of these auxiliary fields in the far infrared (IR) at small $k\rightarrow 0$ in momentum space. We observe a striking similarity of the obtained structure with   analogous formula for the Berry's connection  previously derived for many CM topologically ordered systems.  This analogy  further supports our claim that the ground state of the Maxwell theory  belongs to a topologically ordered phase.  In our Concluding section  \ref{conclusion} we briefly mention possible    settings  where such unusual topological vacuum features can be experimentally  studied. Furthermore, we in fact shall argue that a topological insulator and the topological Maxwell vacuum (studied in this work), while very different in composition, nevertheless,   behave  very much in the same way at large distances.

  \section{Maxwell theory in two dimensions as topological QFT}\label{topology}
  The 2d Maxwell model has been solved numerous number of times using very different techniques, see e.g.\cite{Manton:1985jm,Balachandran:1994vi, SW}. It is known that this is an ``empty" theory in a sense that  it does not support  any propagating degrees of freedom in the bulk of space-time.
  It is also  known that this model can be treated as a conventional topological quantum field theory (TQFT). In particular, this model  can be formulated in terms of     the so-called ``BF"  action involving no metric.   Furthermore, this model   exhibits  many other features such as fractional edge observables which are typical for TQFT, see e.g.\cite{Balachandran:1994vi}. We emphasize on these  properties 
   of the 2d Maxwell theory because the topological portion of the partition function ${\cal{Z}}_{\rm top}$ in our description of 4d Maxwell system, given  in  section \ref{4d},  identically  the same   as the partition function of  2d Maxwell system. Such a relation between the two different systems 
     is a result of  decoupling of physical propagating photons from the topological sectors  in 4d system. 
 
  Our goal here is to review this ``empty" 2d Maxwell theory with nontrivial dynamics of the topological sectors
  when  conventional propagating degrees of freedom are not supported by this system.
    
    \subsection{Partition function and $\theta$ vacua in 2d Maxwell theory}
       
   We consider 2d Maxwell theory defined on the Euclidean torus $S^1\times S^1$ with lengths $L$ and $\beta$ respectively. In the Hamiltonian  framework we choose a $A_0=0$ gauge along with $\partial_1 A_1=0$. This implies that $A_1(t)$ is the only dynamical variable of the system with $E=\dot{A_1}$.
    The spectrum for $\theta$ vacua    is well known \cite{Manton:1985jm}  and it is given by $E_n(\theta)= \frac{1}{2}\left(n+\frac{\theta}{2\pi}\right)^2e^2 L$, such that 
  the corresponding  partition function  takes the form
  \be
 \label{Z_2}
 {\cal{Z}}(V, \theta)=   \sum_{n\in \mathbb{Z}} e^{-\frac{e^2V}{2} \left(n+\frac{\theta}{2\pi}\right)^2},
 \ee
 where $V = \beta L$ is the two-volume of the system.

  We want to reproduce (\ref{Z_2}) using a different approach based on Euclidean path integral computations because it 
  can be easily generalized to similar  computations 4d Maxwell theory defined on 4 torus. Our goal here is to understand the physical meaning of  (\ref{Z_2})  in terms of the path integral computations.

  To proceed with path integral  computations one  considers the  ``instanton"
      configurations   on two dimensional Euclidean torus   with total area $V=L\beta$  described as follows \cite{SW}:
\be
\label{Q}
\int  \dd^2x ~Q(x)  =k,  ~~~~e E^{(k)}=\frac{2\pi k}{V}, 
\ee
 where $Q=\frac{e}{2\pi}E$ is the topological charge density 
and  $k$
is the integer-valued topological charge   in the 2d $U(1)$ gauge theory, $E(x)=\partial_0A_1-\partial_1A_0$ is the field strength. 
The action of this classical configuration is
 \be
 \label{action}
 \frac{1}{2}\int d^2x E^2= \frac{2\pi^2 k^2}{e^2 V}.
 \ee
   This configuration corresponds to the   topological charge $k$ as defined by (\ref{Q}).
The next step is to   compute the  partition function defined as follows
\be
\label{Z_3}
{\cal{Z}}(\theta)=\sum_{ k \in \mathbb{Z}}{\int {\cal{D}}}A^{(k)} {e^{-\frac{1}{2}\int d^2x E^2+ \int d^2x L_{\theta} }},
\ee
where $\theta$ is standard theta parameter which defines the $| \theta\ra$ ground  state and which enters the action with topological density operator
\be
\label{theta}
L_{\theta}=i\theta \int  \dd^2x ~Q(x) =i \theta\frac{e}{2\pi}\int \dd^2x~ E(x). 
\ee
   All integrals in this partition function are gaussian and can be easily evaluated using the technique developed in \cite{SW}. The result is
    \be
 \label{Z_4}
 {\cal{Z}}(V, \theta)= \sqrt{\frac{2\pi}{e^2V}}\sum_{k\in \mathbb{Z}} e^{-\frac{2\pi^2k^2}{e^2V} +ik\theta},
    \ee
    where the expression in the exponent represents the classical instanton configurations with action (\ref{action}) and topological charge (\ref{Q}), while the factor in front is due to the fluctuations, see \cite{Cao:2013na,Zhitnitsky:2013hba} 
    with some technical details and relevant references. 
      While expressions (\ref{Z_2}) and (\ref{Z_4}) look differently, they are actually identically the same, as the Poisson summation formula states:
     \be
 \label{poisson}
  {\cal{Z}}(\theta)= \sum_{n\in \mathbb{Z}} e^{-\frac{e^2V}{2} \left(n+\frac{\theta}{2\pi}\right)^2}= \sqrt{\frac{2\pi}{e^2V}}\sum_{k\in \mathbb{Z}} e^{-\frac{2\pi^2k^2}{e^2V} +ik\theta}.~~~~
    \ee
   Therefore, we reproduce the original expression (\ref{Z_2})  using the path integral approach. 
   
    The crucial observation for our present study is that this naively ``empty" theory which has no physical propagating degrees of freedom, nevertheless shows some very nontrivial features of the ground state related to the topological properties of the   theory.  These new properties are formulated in terms  of different topological vacuum sectors 
of the system $|k\ra$ which have identical physical properties as they connected to each other by  large gauge transformation operator ${\cal T}$
 commuting   with the Hamiltonian $[{\cal T}, H]=0$. As explained in details in  \cite{Cao:2013na,Zhitnitsky:2013hba}  the corresponding dynamics of this ``empty" theory represented by partition function (\ref{poisson}) should be interpreted as a result of tunnelling events between  these ``degenerate" winding $ | k\ra$ states which correspond to one and the same physical state.

It is known that this model can be treated as TQFT, e.g. supports edge observables which may assume the  fractional values, and shows many other features which are typical for a TQFT, see \cite{Balachandran:1994vi} and references therein.  The presence of the  topological features of the model can be easily understood from  observation  that entire dynamics of the system is due to the transitions between the topological sectors which themselves are determined by the behaviour of surface integrals at infinity $\oint A_{\mu}dx^{\mu}$. These sectors are classified by   integer numbers 
and they are not sensitive to specific details of the system such as geometrical shape of the system. Therefore, it is not really a surprise that the system is not sensitive to specific geometrical details, and can be treated as TQFT. 
The simplest way to analyze  the corresponding topological features of the system is to introduce the topological susceptibility $\chi$ and study its property, see next subsection.

    \subsection{Topological susceptibility}\label{top_2}
The topological susceptibility $\chi $  is defined as follows,
  \be
\label{chi1}
\chi \equiv   \lim_{k\rightarrow 0} \int \dd^2x ~e^{ikx}\left< T Q(x) Q(0) \right> ,
\ee
where $Q $ is topological charge density operator normalized according to  eq.(\ref{Q}).  
The   $\chi$  measures response  of the free energy to the introduction of a source  term  
 defined by eq.  (\ref{theta}).
The   computations of $\chi$ in this simple ``empty" model  can be easily carried out  as the partition function ${\cal{Z}}(\theta)$ defined by (\ref{Z_3}) is known exactly (\ref{poisson}). To compute $\chi$ we  should simply differentiate the partition function twice with respect to $\theta$. It leads  to the following well known expression for $\chi$ 
 which is finite  in the infinite volume limit \cite{SW,Zhitnitsky:2011tr,Zhitnitsky:2013hba}
\be
\label{exact1}
\chi (V\rightarrow \infty)= -\frac{1}{V }\cdot\frac{\partial^2 {\ln\cal{Z}}(\theta)}{\partial\theta^2}|_{\theta=0}= \frac{e^2}{4\pi^2}.  
\ee 
A typical value of the topological charge $k$ which saturates the topological susceptibility $\chi$ in the large volume limit is very large, $k\sim \sqrt{e^2 V}\gg 1$.

  It is important to emphasize that the integrand for the topological susceptibility (\ref{chi1})   demonstrates a singular behaviour, see \cite{SW,Zhitnitsky:2011tr,Zhitnitsky:2013hba}
  for the details and related references:
 \be
   \label{local}
   \left<  Q(x) Q(0) \right>= \frac{e^2}{4\pi^2}\delta^2 (x).
   \ee
It represents the non-dispersive contact term which can  not be related to any propagating degrees of freedom.
In this simplest case of the  2d Maxwell system this comment is quite obvious as 2d Maxwell theory does not support any propagating degrees of freedom. 
   The $\delta^2 (x)$ function in  (\ref{local}) should be understood as total divergence related to the infrared (IR) physics, rather than to ultraviolet (UV) behaviour. Indeed,  
\be	\label{divergence}
  \chi  &=& \frac{e^2}{4\pi^2}  \int\delta^2 (x) \dd^2x  
=   \frac{e^2}{4\pi^2}\int   \dd^2x~
  \partial_{\mu}\left(\frac{x^{\mu}}{2\pi x^2}\right)\nonumber\\
  &=&  \frac{e^2}{4\pi^2} \oint_{S_1\rightarrow \infty}    \dd l_{\mu}\epsilon^{\mu\nu}
 \left(\frac{x_{\nu}}{2\pi x^2}\right) =   \frac{e^2}{4\pi^2}.
\ee
  In other words, the non-dispersive contact term   (\ref{local}) is  determined by  IR physics
 at arbitrary large distances   rather than UV physics which can be erroneously assumed to be  a source of   
 $\delta^2 (x)$ behaviour in (\ref{local}). 
 The  computations of this contact term in terms of the delocalized instantons (\ref{Q}) explicitly  show that   all observables in this system are originated from the IR physics. 
   
One should also  remark  that   the same contact  term (\ref{exact1}) and its local expression (\ref{local}) can be also computed  using the auxiliary ghost field, the so-called Kogut-Susskind (KS) ghost,  as it has been originally done in ref. \cite{KS}, see also \cite{Zhitnitsky:2011tr,Zhitnitsky:2013hba} for relevant  discussions in the present context.    This   description in terms of the KS ghost  implicitly  takes into account the presence  of  topological sectors in the system. The same property  is explicitly reflected   by summation over topological sectors $ { k \in \mathbb{Z}}$ in direct  computations  (\ref{Z_3},\ref{Z_4}) without introducing any auxiliary fields. 
 
    \exclude{
    Important  point we would like to make is that our analysis of the topological portion $ {\cal{Z}}_{\rm top}$ of the partition function for   4d Maxwell system defined on ${\mathbb{T}}^4$ assumes exactly the same form (\ref{poisson}) as a result of decoupling of propagating photons from the topological part of the partition function,  as will be discussed in section \ref{4d}. As a result of  this decoupling the topological portion of the 4d Maxwell system behaves very much in the same way as  2d ``empty" theory. Therefore, one should not be very surprised that this 4d system also 
demonstrates some topological features, similar to 2d system reviewed in this section.
}

\section{Topological partition function in 4d}\label{4d}
 Our goal here is to analyze  the Maxwell system   on a Euclidean 4-torus   with  sizes $L_1 \times L_2 \times L_3 \times \beta$ in the respective directions. It provides the infrared (IR) regularization of the system. 
 This IR regularization plays a key role in proper treatment of the  topological terms which are  related to tunnelling events 
 between topologically distinct but physically identical states. 
First, we want to review   the  previously known results on the vacuum structure of this system.  As the second step, we want to  reproduce   these known results on Maxwell vacuum state using a different  technique based on the auxiliary fields to be developed in next section  \ref{berry-section}. As we argue  in  section \ref{4d_auxiliary}   precisely  these    auxiliary topological fields  have  exactly the same mathematical properties  as  emergent Berry's connection in  topologically ordered CM systems.   
 
 \subsection{Construction}
 We follow \cite{Cao:2013na,Zhitnitsky:2013hba} in our construction of the partition function ${\cal{Z}}_{\rm top}$ where it was employed    for  computation of  the corrections to the Casimir effect due to these novel type of topological fluctuations. The crucial point is that we impose the periodic boundary conditions on gauge $A^{\mu}$ field up to a large gauge transformation.
 In what follows we simplify our analysis by considering   a clear case with winding topological sectors $|k\ra$    in the z-direction only.  The classical configuration in Euclidean space  which describes the corresponding tunnelling transitions can be represented as follows:
\be
\label{topB4d}
\vec{B}_{\rm top} &=& \vec{\nabla} \times \vec{A}_{\rm top} = \left(0 ,~ 0,~ \frac{2 \pi k}{e L_{1} L_{2}} \right),\\
\Phi&=&e\int dx_1dx_2  {B}_{\rm top}^z={2\pi}k \nonumber
\ee
in close analogy with the 2d case (\ref{Q}).

The Euclidean action of the system is quadratic and has the following  form  
\be
\label{action4d}
\frac{1}{2} \int \dd^4 x \left\{  \vec{E}^2 +  \left(\vec{B} + \vec{B}_{\rm top}\right)^2 \right\} ,
\ee
where $\vec{E}$ and $\vec{B}$ are the dynamical quantum fluctuations of the gauge field.  
The key point is that the classical topological portion of the action decouples from quantum fluctuations, such that the quantum fluctuations do not depend on topological sector $k$ and can be computed in topologically trivial sector $k=0$.
Indeed,  the cross term 
\be
\int \dd^4 x~ \vec{B} \cdot \vec{B}_{\rm top} = \frac{2 \pi k}{e L_{1} L_{2}} \int \dd^4 x~ B_{z} = 0 
\label{decouple}
\ee
vanishes  because the magnetic portion of quantum fluctuations in the $z$-direction, represented by $B_{z} = \partial_{x} A_{y}  - \partial_{y} A_{x} $, is a periodic function as   $\vec{A} $ is periodic over the domain of integration. 
This technical remark in fact greatly simplifies our  analysis as the contribution of the physical propagating photons 
is not sensitive to the topological sectors $k$. This is,  of course,  a specific feature  of quadratic action 
 (\ref{action4d}), in contrast with non-abelian  and non-linear gauge field theories where quantum fluctuations of course depend on topological $k$ sectors. The authors of 
 ref. \cite{ChenLee} arrived to the same   conclusion (on decoupling  of the  topological terms from  conventional fluctuating photons with non-zero momentum),   though in a different context of topological insulators in the presence of the $\theta=\pi$ term. 
 
The classical action  for configuration (\ref{topB4d}) takes the form 
\be
\label{action4d2}
\frac{1}{2}\int \dd^4 x \vec{B}_{\rm top}^2= \frac{2\pi^2 k^2 \beta L_3}{e^2 L_1 L_2}
\ee
To simplify our analysis further in  computing  ${\cal{Z}}_{\rm top}$ we consider a geometry where $L_1, L_2 \gg L_3 , \beta$ similar to construction relevant for the Casimir effect  \cite{Cao:2013na,Zhitnitsky:2013hba}. 
 In this case our system   is closely related to 2d Maxwell theory by dimensional reduction: taking a slice of the 4d system in the $xy$-plane will yield precisely the topological features of the 2d torus considered in section \ref{topology}.  
 Furthermore, with this geometry our simplification (\ref{topB4d}) when we consider exclusively the magnetic fluxes in $z$ direction is justified as the corresponding classical action (\ref{action4d2}) assumes a minimal  possible values.  With this assumption we can consider very small temperature, but still we can not take a formal limit $\beta\rightarrow\infty$  in our final expressions
 as a result of our technical constraints in the system. 
      
With these additional simplifications   the topological partition function becomes \cite{Cao:2013na,Zhitnitsky:2013hba}:
\be
\label{Z4d}
{\cal{Z}}_{\rm top} = \sqrt{\frac{2\pi \beta L_3}{e^2 L_1 L_2}} \sum_{k\in \mathbb{Z}} e^{-\frac{2\pi^2 k^2 \beta L_3}{e^2 L_1 L_2} }= \sqrt{\pi \tau} \sum_{k\in \mathbb{Z}}e^{-\pi^2 \tau k^2}, ~~~~
\ee
where we introduced the dimensionless parameter
\be
\label{tau}
\tau \equiv {2 \beta L_3}/{e^2 L_1 L_2}.
\ee
Formula (\ref{Z4d})   is essentially the dimensionally reduced expression for  the topological partition function (\ref{Z_4}) for 2d 
 Maxwell theory analyzed in section \ref{topology}. 
   One should note that the normalization factor $\sqrt{\pi \tau}$ which appears in eq. (\ref{Z4d}) does not depend on topological sector $k$, and essentially it represents our convention of the  normalization   ${\cal{Z}}_{\rm top}
\rightarrow 1$ in the limit $L_1L_2\rightarrow \infty$ which corresponds to  a convenient set up for the  Casimir -type experiments as discussed in \cite{Cao:2013na,Zhitnitsky:2013hba}.

\subsection{ External magnetic field }\label{magnetic}

In this section we want to generalize our results   for the Euclidean Maxwell system in the presence of the external magnetic field. Normally, in the conventional quantization of electromagnetic fields in infinite Minkowski space, there is no \emph{direct} coupling    between fluctuating vacuum photons and an external magnetic field as a consequence of linearity of the Maxwell system. The coupling with  fermions   generates  a negligible effect $\sim \alpha^2B_{ext}^2/m_e^4$ as the non-linear Euler-Heisenberg Effective Lagrangian  suggests, see \cite{Cao:2013na} for the details and numerical estimates.  The interaction of the external magnetic field with topological fluctuations (\ref{topB4d}),  in contrast with coupling with conventional   photons, will  lead to the effects of order of unity as a result of interference of the external magnetic field with topological fluxes $k$.

The corresponding partition function can be easily constructed for external magnetic field $B_{z}^{\rm ext}$ pointing along $z$ direction, as 
the crucial technical element on decoupling of the background fields  from quantum fluctuations assumes the same form (\ref{decouple}).  In other words, the physical propagating photons with non-vanishing momenta are not sensitive to the topological $k$ sectors, nor to the external uniform magnetic field, similar to our discussions after (\ref{decouple}).

The classical action for configuration in the presence of the uniform external magnetic field $B_{z}^{\rm ext}$ therefore takes the form 
\be
\label{B_ext}
\frac{1}{2}\int \dd^4 x  \left(\vec{B}_{\rm ext} + \vec{B}_{\rm top}\right)^2=  \pi^2\tau\left(k+\frac{\theta_{\rm eff}}{2\pi} \right)^2
\ee
where $\tau$ is defined by (\ref{tau}) and  the effective theta parameter $\theta_{\rm eff} \equiv eL_1L_2 B^z_{\rm ext}$ is expressed in terms of the original external magnetic field $B^z_{\rm ext}$.
Therefore, the partition function in the presence of the uniform magnetic field can be easily reconstructed from (\ref{Z4d}), and it is given by \cite{Cao:2013na,Zhitnitsky:2013hba}
\be 
\label{Z_eff}
  {\cal{Z}}_{\rm top}(\tau, \theta_{\rm eff})
 =\sqrt{\pi\tau} \sum_{k \in \mathbb{Z}} \exp\left[-\pi^2\tau \left(k+\frac{\theta_{\rm eff}}{2\pi}\right)^2\right].~~
\ee
This system in what follows will be referred as the topological vacuum (  $\cal{TV}$) because the  propagating degrees of freedom, the photons with two transverse polarizations,   
completely decouple from  ${\cal{Z}}_{\rm top}(\tau, \theta_{\rm eff})$.

  The dual representation for the partition function is obtained by applying the Poisson summation formula (\ref{poisson}) such that (\ref{Z_eff}) becomes 
  \be 
\label{Z_dual1}
  {\cal{Z}}_{\rm top}(\tau, \theta_{\rm eff})
  = \sum_{n\in \mathbb{Z}} \exp\left[-\frac{n^2}{\tau}+in\cdot\theta_{\rm eff}\right]. 
  \ee
 Formula (\ref{Z_dual1})  justifies our notation for  the effective theta parameter $\theta_{\rm eff}$ as it enters the partition function in combination with integer number $n$. One should emphasize that integer  number $n$ in the dual representation (\ref{Z_dual1}) is not the integer magnetic flux $k$ defined by eq. 
(\ref{topB4d}) which enters the original partition function (\ref{Z4d}). Furthermore,  the $\theta_{\rm eff}$ parameter which enters (\ref{Z_eff}, \ref{Z_dual1}) is not a fundamental $\theta$ parameter which is normally introduced into the Lagrangian  in front of  $\vec{E}\cdot\vec{B}$ operator. Rather, this parameter  $\theta_{\rm eff}$ should be understood as an effective parameter representing the construction of the  $\theta_{\rm eff}$ state for each slice in four dimensional system. In fact, there are three such  $\theta_{\rm eff}$  parameters representing different slices and corresponding external magnetic fluxes. There are similar three $\theta_i$ 
parameters representing the external electric fluxes   as discussed  in \cite{Zhitnitsky:2013hba}, such that total number of $\theta$ parameters classifying the system equals six, in agreement with total number of hyperplanes in four dimensions. 

\section{Berry Connection}\label{berry-section}
The main goal of this section is to argue that our   $\cal{TV}$- configuration represents a simplest version of a topologically ordered phase very similar to CM systems  \cite{Wen:1989iv,Wen:1990zza,Moore:1991ks,BF,Cho:2010rk,Wen:2012hm,Sachdev:2012dq, Cortijo:2011aa, Volovik:2011kg}. 
We want to reformulate the topological features of the system   (analyzed   in  section \ref{4d}) in terms of the Berry's connection and Berry curvature normally computed in momentum space in CM literature. Such a deep relation between the two very different descriptions will demonstrate once again that the ground state for the Maxwell theory defined on a compact manifold exhibits all the features which are normally attributed to  a topologically ordered system. 
We make this relation  much more precise  by introducing the auxiliary topological fields which can be identified with Berry's connection. With such an interpretation the complex phase in the dual representation  (\ref{Z_dual1})  can be thought as  the Berry's phase  which is known to emerge in many quantum systems. 

We start our study in section \ref{2d-berry}  by reviewing the well-known CM results  on the Berry's connection. In section \ref{2d_auxiliary} 
we describe  the ground state of the two dimensional Maxwell theory by  using the auxiliary topological fields.  
We observe a deep mathematical similarity between the Berry's connection computed for CM systems (including the monopole-type behaviour in momentum $\mathbf{k}$ space) and the corresponding formulae computed for the ground state in the Maxwell theory in terms of the auxiliary topological fields. 
We generalize the corresponding construction      to four dimensional Maxwell system defined on a four -torus in section \ref{4d_auxiliary}.

\subsection{Berry phase in CM  systems}\label{2d-berry} 

In this subsection we review the computations of the Berry connection in some CM systems. 
 In context of the topological insulators and quantum Hall systems the corresponding studies have been carried out  in  two,   three and four dimensions \cite{Wen:1989iv,Wen:1990zza,Moore:1991ks,BF,Cho:2010rk,Wen:2012hm,Sachdev:2012dq, Cortijo:2011aa, Volovik:2011kg}, see also \cite{Zhitnitsky:2013hba,Thacker:2013nma} with related discussions of the  ground state in 2d Maxwell theory\footnote{Not to be confused with conventional CM notations, where it is a customary to count the spatial number of dimensions, rather than total number of dimensions.  For our 2d system this convention corresponds to (D + 1) Maxwell theory with D = 1. Similar studies have been carried out   for topological insulators  for D=1 and    D=3, see  e.g. \cite{ChenLee} with many  references on the original results.
For   D=2 the corresponding computations  of the Berry's connection for the  quantum Hall systems have been reviewed in \cite{Cortijo:2011aa}.}. 

In the simplest D=1 case the expression for the Berry's phase (which is the accumulated geometric phase of the band electrons under the process when the winding of the gauge field is increased by one unit) can be computed as follows\cite{ChenLee}. In the physical $A_0=0$ gauge it corresponds to a slow variation of gauge filed $eA_1$ from $\frac{2\pi n}{L}$ to $\frac{2\pi (n+1)}{L}$ where $L$ is the size of a torus along $x$ direction. The relevant formula is given by \cite{ChenLee}
\be
\label{berry}
\phi_{\rm Berry}=i\int dA_1\la \Psi_{\theta} |\frac{\partial}{\partial A_1}|\Psi_{\theta} \ra ,
\ee
where $|\Psi_{\theta} \ra$ is the full wave function of the system which can be   expressed in terms of single particle wave functions. One can explicitly demonstrate  \cite{ChenLee} that  $\phi_{\rm Berry}=-2\pi P$ with $P$ being the polarization of the system  such that $\theta$ is shifted as follows $\theta\rightarrow (\theta -2\pi P)$. The key observation in this computation is that the integration over slow varying gauge fields in eq. (\ref{berry}) is reduced to integration over allowed momentum $k$ covering the whole Brillouin zone (BZ), i.e.
  \be
\label{berry1}
\phi_{\rm Berry}=i\int_{BZ} dk\la \Psi_{\theta} |\frac{\partial}{\partial k}|\Psi_{\theta} \ra \equiv \int_{BZ} dk {\cal{A}}(k), 
\ee
where ${\cal{A}}(k)$ is the so-called Berry's connection in the momentum space. 
A simple technical explanation of this  key technical  step   (related to the  change of   variables) is  that the large gauge transformation formulated in terms of $A_1$ can be expressed  in terms of  a shift of the  momentum $k$  when the system returns to   the  physically identical  (but topologically different)  state.

Similar computations can be also carried out  for integer quantum Hall system for D=2, in which case the corresponding formula
for the Berry's connection and Berry's curvature takes the form, see e.g. \cite{Cortijo:2011aa}:
\be
\label{berry2}
{\cal{A}}^j(\mathbf{k})=\frac{\tau}{2}\frac{\epsilon^{ij}k_i}{\mathbf{k}^2}, ~~~ {\cal{B}}(\mathbf{k})=\frac{\tau}{2}\delta^2(\mathbf{k}),
\ee
where $\tau=\pm 1$ describe the degenerate Fermi points with the linear dispersion relation $\epsilon (\mathbf{k})\sim |\mathbf{k}|$.
One can identify the behaviour (\ref{berry2}) with magnetic monopoles in momentum space with half-integer magnetic charges. As we shall see below in section  \ref{2d_auxiliary}  a very similar structure also emerges in description of the ground state of the 2d Maxwell system, when the auxiliary topological fields play the role of the  Berry's connection (\ref{berry2}).

One should emphasize that in CM literature the corresponding ${\cal{A}}^j(\mathbf{k})$   fields are the emergent gauge fields. The real   source for these emergent gauge configurations is the  strongly coupled coherent superposition  of the  physical  electrons. In contrast, in our case  a formula to be derived below (and which mathematically identical    to eq. (\ref{berry2})) will arise  from   the  topologically non-trivial gauge configurations of the  underlying fundamental  gauge theory. In other words, in our case the  formula similar to  (\ref{berry2})   will emerge as  a result of the  topologically non-trivial vacuum gauge configurations which are present in the system irrespectively to existence  of the fermions. 

In the following subsection \ref{2d_auxiliary} we reformulate the  known results about the ground $\theta$ state in 2d Maxwell system  
 using  the    topological  auxiliary  (non-propagating) fields. The corresponding technique, as we shall see below in section \ref{4d_auxiliary}, can be easily generalized to   four dimensional  Maxwell system, which is the main subject of the present work. 

 \subsection{Auxiliary topological fields in 2d Maxwell theory}\label{2d_auxiliary}
 We wish to derive the topological action for the Maxwell system in 2d by using a standard conventional  technique exploited e.g.   in  \cite{BF} for the Higgs model in CM context or  in \cite{Zhitnitsky:2013hs} for the so-called weakly coupled ``deformed QCD". 
 We shall reproduce below the well- known results for this ``empty" 2d system   including a  non-vanishing
 expression for the topological susceptibility (\ref{exact1},\ref{local}) using the corresponding auxiliary fields in momentum space. 
 It turns out that the corresponding connection and curvature computed using these auxiliary fields play the same role as the Berry's connection and Berry's curvature play in CM systems. 
  To be more precise,  the unique topological features of the auxiliary field is precisely the key element which allows to represent the accumulated geometric phase in terms of the auxiliary field  sensitive to the boundary conditions. An explicit demonstration of such a relation  between the Berry's phase and auxiliary topological fields     is precisely the main subject of this section.  
 
 Our starting point is to  insert  the  delta function  into the path  integral with the field $b(\mathbf{x})$ acting as a Lagrange multiplier
 \be
 \label{delta}
 &&\delta \left[Q(\mathbf{x})-\frac{e}{2\pi}\epsilon^{jk} \partial_{j} a_k(\mathbf{x})\right]\sim \\
&& \int {\cal{D}}[b] e^{ i \int d^2\mathbf{x} ~b(\mathbf{x})\cdot  \left[ Q(\mathbf{x})-\frac{e}{2\pi}\epsilon^{jk} \partial_{j} a_k(\mathbf{x})\right]} \nonumber
  \ee
 where $Q(\mathbf{x})=\frac{e}{2\pi}E(\mathbf{x})$ in this formula  is the topological charge density operator. It will be       treated as the original expression for the field operator entering the action (\ref{Z_3}) with topological term (\ref{theta}).
 At the same time  $a_k(\mathbf{x})$  is treated as a slow-varying   external source effectively describing the large distance physics  for a given instanton configuration. 
The insertion (\ref{delta}) of  the  delta function assumes that the path integral computations must include  all  the classical k-instanton configurations (\ref{Q}),(\ref{action})  along with  quantum fluctuations surrounding them. In other words, we treat  $Q(\mathbf{x})$ as a fast degree of freedom, while $a_k(\mathbf{x})$ are considered as    slow degrees of freedom representing an external background field.  

One should remark here that the corresponding formal manipulation is not a mathematically  rigorous procedure 
as  $a_k(\mathbf{x})$  must be  singular somewhere to support non-vanishing topological charges in the system\footnote{I am thankful to anonymous referee for pointing this out.}. 
The presence of such singularity is very similar to emergent singularities  in description of the Berry's connection, Dirac's string, or  the Aharonov Bohm  potential.   
It is not a goal of the present work to search for a more  rigorous mathematical tools for corresponding problems. The most important argument for us that our procedure  represented by eq.(\ref{delta})  is correct is the fact that  the topological susceptibility (\ref{top2}), (\ref{final}) as well as the expectation value of the electric field 
(\ref{Q1}), (\ref{E}) are precisely reproduced when computations are performed  with  our formal approach    utilizing the  auxiliary topological fields. 

Another point worth to be mentioned is as follows. As we stated above, the auxiliary field $a_k(\mathbf{x})$ is treated as a slow field, while  $Q(\mathbf{x})$ is treated as a fast degree of freedom. At the same time,   formally, these fields are proportional to each other $Q(\mathbf{x})\sim \epsilon^{jk} \partial_{j} a_k(\mathbf{x})$
according to (\ref{delta}), and therefore, it is not obvious how   these fields  could be  treated so differently.
The answer lies in the observation that our  auxiliary fields   $a_k(\mathbf{x}), b_z(\mathbf{x})$ are non-dynamical fields, have no kinetic terms, and     do not propagate,   in contrast with conventional gauge fields.  
Formally, these fields do not have their conjugate momenta, as they are auxiliary non-dynamical fields of the system. 

The simplest way to understand  this construction   is through analogy with well known and well understood model  in particles physics, the so-called  Nambu-Jona-Lasino model. In this case an  auxiliary $\sigma$ field without kinetic term is introduced into the system, analogous to (\ref{delta}).  The $\sigma\sim <\bar{\psi}\psi>$ field is treated as a slow field and  in mean field approximation represents the chiral condensate of the fermi-fields. Our auxiliary fields   $a_k(\mathbf{x}), b_z(\mathbf{x})$ should be understood exactly  in the same way as $\sigma$ field is is understood in  Nambu-Jona-Lasino model.

Now we are coming back to our proposed formula (\ref{delta}). Our task now is to integrate out the original   fast ``instantons"  and describe the large distance physics in terms of slow varying fields $b(\mathbf{x}), a_k(\mathbf{x})$ in form of the effective action $S_{\rm top}[b, a_k] $ formulated in terms of slow auxiliary fields$b(\mathbf{x}), a_k(\mathbf{x})$. We use conventional well established procedure of summation over k-instantons 
reviewed in section \ref{topology} with final result (\ref{Z_4}).  The only new element 
in comparison with the previous computations is that  the fast degrees of freedom must be integrated out in the presence of the new slow varying  background fields   $b(\mathbf{x}), a_k(\mathbf{x})$ which appear in  eq. (\ref{delta}). Fortunately, the computations can be easily performed for such external sources. Indeed,  one should notice  that the background field $b(\mathbf{x})$ enters eq. (\ref{delta}) exactly in the same manner as external parameter $\theta$    enters  (\ref{theta}).
Therefore, assuming that $b(\mathbf{x}), a_k(\mathbf{x})$ are slow varying background fields we arrive to the following 
  expression for the   partition function:
  \be
\label{Z_BF}
{\cal{Z}}_{\rm top}  =  \int {\cal{D}}[b]  {\cal{D}}[a]  e^{-\frac{e^2}{8\pi^2} \cdot \int d^2 \mathbf{x}
\left[ \theta+b(\mathbf{x})\right]^2-S_{\rm top}},~~
\ee
where $  b (\mathbf{x})$  represents the slow varying background auxiliary   field
which is assumed to lie in the lowest $n=0$ branch,    $|b(\mathbf{x})|< \pi $.  Correspondingly, 
in formula (\ref{Z_BF})   we kept only the asymptotically leading term in expansion (\ref{Z_2})   with $n=0$  in   large volume limit, $(e^2V)\gg 1$.  
The topological term  $S_{\rm top}[b, a_k] $ in eq. (\ref{Z_BF}) reads  
\be
\label{S_top}
 S_{\rm top}[b, a_k] 
 =  i  \frac{e}{2\pi} \int d^2\mathbf{x} \left[b(\mathbf{x}) \epsilon^{jk} \partial_{j} a_k(\mathbf{x})\right].~~~~~
\ee
 
  Our goal now  is to consider a simplest  application of the effective low energy topological action   (\ref{Z_BF}), (\ref{S_top}) we just derived. We want   to reproduce the known  expression for  the   topological   susceptibility (\ref{exact1}),(\ref{local}) by integrating out the $b$ and $a_k$ fields  using low energy effective description  (\ref{Z_BF}), (\ref{S_top}), rather than an explicit summation over the instantons, which was employed in the original derivation (\ref{exact1}),(\ref{local}). The agreement between of two drastically different approaches   will give us a confidence that our formal manipulations with the auxiliary fields is a correct and self-consistent procedure. With this confidence, as a next step,  we will study the behaviour of  the auxiliary topological fields  in the IR, which corresponds to $k\rightarrow 0$ in momentum space.  We  compare the corresponding formula  with the Berry's connection at small $k\rightarrow 0$ to observe that both expressions behave in a very similar way at large distances  in the IR. Such a similarity allows us to identify the auxiliary field $a_k(\mathbf{x})$  governed by the action  (\ref{Z_BF}), (\ref{S_top}) with emergent Berry's connection ${\cal{A}}^j(\mathbf{k})$ given by eq.(\ref{berry2}).
  
  To proceed with this task we compute the topological susceptibility at $\theta=0$ as follows, 
  \be
  \label{top1}
  \left<  Q(\mathbf{x}) Q(\mathbf{0}) \right> &=&\frac{1}{\cal{Z}}  \int {\cal{D}}[b]  {\cal{D}}[a]  e^{-S_{\rm tot}[b, a_k] } \nonumber\\
  \frac{e^2}{4\pi^2}&\cdot&\left[ \epsilon^{jk} \partial_{j} a_k(\mathbf{x}) ,\epsilon^{j'k'} \partial_{j'} a_{k'}(\mathbf{0})\right], 
  \ee
 where $S_{\rm tot}[b, a_k] $ determines the dynamics of  auxiliary $b$ and  $a_k$ fields, and  it is given by
 \be
 \label{S_tot}
S_{\rm tot}[b, a_k]= \int d^2\mathbf{x}\left[ \frac{e^2}{8\pi^2} b^2(\mathbf{x})
 +i \frac{e}{2\pi}b(\mathbf{x}) \epsilon^{jk} \partial_{j} a_k(\mathbf{x})\right].~~~~~
 \ee
 The obtained   Gaussian integral (\ref{top1}) over $\int {\cal{D}}[b]$ can be explicitly executed, and we are left with the following   integral over $\int {\cal{D}}[a]$
  \be
  \label{top2}
  &&  \left<  Q(\mathbf{x}) Q(\mathbf{0}) \right> =  \frac{1}{\cal{Z}}  \int  {\cal{D}}[a]e^{- \frac{1}{2}\int d^2\mathbf{x} \left[ \epsilon^{jk} \partial_{j} a_k(\mathbf{x})\right]^2} \nonumber\\
 &\cdot&    \frac{e^2}{4\pi^2}  \left[ \epsilon^{jk} \partial_{j} a_k(\mathbf{x}) ,\epsilon^{j'k'} \partial_{j'} a_{k'}(\mathbf{0})\right].  
  \ee
 The    integral (\ref{top2}) is also gaussian and can be explicitly evaluated  with the following final result
 \be
 \label{final}
   && \left<  Q(\mathbf{x}) Q(\mathbf{0}) \right> =  \frac{e^2}{4\pi^2} \delta^2 (\mathbf{x}), \\
    &&  \int d^2\mathbf{x} \left<  Q(\mathbf{x}) Q(\mathbf{0}) \right> =   \frac{e^2}{4\pi^2}. \nonumber
 \ee
   Few comments are in order. First, formula (\ref{final})  precisely reproduces our previous  expression (\ref{exact1}),(\ref{local}) derived by explicit summation over fluxes-instantons, and without even mentioning any auxiliary topological fields 
 $b (\mathbf{x}), a_k (\mathbf{x})$. It obviously demonstrates  a self-consistency of our formal manipulations with auxiliary topological fields. As we shall see below, the reformulation of the system in terms of the auxiliary topological fields is extremely useful for studying some other (very non-trivial) topological features of the gauge system.
 
 Secondly, the expression (\ref{final}) for the topological  susceptibility represents the contact non-dispersive term   which can not be associated with any physical propagating degrees of freedom as we discussed in section \ref{top_2}. The nature of this  contact term can be understood in terms of   the  tunnelling transitions between topologically different but physically identical $|k\ra$ states. As we already mentioned in section \ref{top_2} the same  contact term   can be also understood in terms of the propagating  Kogut-Susskind ghost\cite{KS}, 
 which effectively describes  the  tunnelling transitions in terms of an auxiliary Kogut-Susskind ghost
 which however, does not belong to the physical Hilbert space, see \cite{Zhitnitsky:2011tr} for the  details in given context.

To proceed with our task on establishing the  relation between  the topological auxiliary fields and  the Berry's connection we want compute the expectation value for the topological charge density operator $\left< Q \right>  \equiv   \la \frac{e }{2\pi}  E \ra$ at non-vanishing $\theta\neq 0$. The corresponding computations can be easily performed  using the same technique described above. The only new element which occurs is necessity to   compute  the path integral   at non-vanishing $\theta$ as the entire final result  will be  proportional to $\theta$, see eq. (\ref{Q1}) below. However, the presence of $\theta$ in the effective action  does not 
produce any technical difficulties  as the emergent path integral    remains to be the Gaussian integral  determined by the  quadratic action (\ref{Z_BF}) even for non-vanishing $\theta$. The corresponding computation at $\theta\neq 0$ can be easily executed by a conventional shift of variables  $b (\mathbf{x}), a_k (\mathbf{x})$.   The result is: 
\be
 \label{Q1}
\lim_{\mathbf{k}\rightarrow 0}&& \int d^2\mathbf{x} e^{i\mathbf{kx}} \left<  Q(\mathbf{x}) \right> \\
=\lim_{\mathbf{k}\rightarrow 0} &&\left(\frac{e}{2\pi}\right)\int d^2\mathbf{x} e^{i\mathbf{kx}}\la\epsilon^{ij} \partial_{i} a_j(\mathbf{x})\ra=
  \frac{ie^2\theta}{4\pi^2} V ,   \nonumber
 \ee
where $V$ is the total volume of the system   playing  the role of the IR regulator  in all computations in 2d Maxwell system as reviewed  in section \ref{topology}. The obtained formula (\ref{Q1}) reproduces the well-known result   that a non-vanishing $\theta$ corresponds to non-vanishing  background  electric field $E   \equiv\epsilon^{ij} \partial_{i} a_j $ in the system \cite{Print-76-0357 (HARVARD)}, 
\be
\label{E}
\la E\ra_{\rm Eucl.}= \frac{ie\theta}{2\pi}, ~~~~\la E\ra_{\rm Mink.}=\frac{e\theta}{2\pi},~~~
\ee
see also \cite{Zhitnitsky:2013hba} with some comments in the given context.

The non-vanishing expectation value of the gauge invariant operator (\ref{Q1}) is highly non-trivial phenomenon as the operator $Q(\mathbf{x}) $ itself is a total divergence. Naively, all correlation functions with operator   $Q(\mathbf{x}) $, including  the expectation value of $\left<  Q(\mathbf{x}) \right> $ itself must  vanish  in $k\rightarrow 0$ limit as there are no any physical massless degrees of freedom in the system. We know that this naive conclusion is incorrect as   well established results  (\ref{chi1}), (\ref{exact1}),  (\ref{Q1}), (\ref{E})    explicitly show. The loophole in  the aforementioned   naive conclusion is related to the generating  of the non-dispersive (contact) contributions which can not be formulated in terms of any physical propagating degrees of freedom. 
\exclude{
The corresponding  contact term in  (\ref{chi1}), (\ref{exact1}) is  known to be saturated by the topological instanton-like configurations  (\ref{Q}) which are  originated from   different topological sectors  of  the system. This contact term does not vanish even in 2d Maxwell theory  (\ref{exact1}) when physical degrees of freedom simply do not exist in the system.  }
The same IR physics, as we already mentioned,  can be also described in terms of the massless Kogut -Susskind ghost \cite{KS} which effectively (implicitly) describes   the   tunnelling transitions between the topological $|k\ra$ sectors\footnote{Furthermore, one can argue that the topological auxiliary field $a_i(\mathbf{k})$ introduced above can be   expressed in terms of the   Kogut -Susskind ghost. Apparently, such a relation is very generic feature of many gauge theories. In fact, an analogous relation can be explicitly worked out    in four dimensional gauge theory, in   the so-called  weakly coupled ``deformed QCD" where the auxiliary topological fields, similar $a_i (\mathbf{x}), b(\mathbf{x}) $ fields from (\ref{S_tot}) are  related to  the Veneziano ghost~\cite{Zhitnitsky:2013hs}.    The Veneziano ghost was postulated in QCD  long ago  \cite{ven}    with the sole purpose  to saturate the non-dispersive (contact) term in  topological susceptibility, similar in structure to eq. (\ref{final}).  As it is known this contact term plays  the key role in the resolution of the so-called $U(1)_A$ problem in QCD ~\cite{ven, witten}.}.

Such a strong IR sensitivity implies that the Fourier transform of the auxiliary   topological field $a_j$ saturating the expectation value (\ref{Q1}) has the  singular behaviour at small 
momentum $k\rightarrow 0$: 
\be
 \label{k}
 a^j(\mathbf{k}\rightarrow 0)\equiv \frac{1}{V}\int d^2\mathbf{x} e^{i\mathbf{kx}} a^j(\mathbf{x})\rightarrow 
\left( \frac{e\theta}{2\pi}\right)\frac{\epsilon^{ij}   {k}_i}{ \mathbf{k}^2}, 
 \ee
 in spite of the fact that the system does not support any physical massless propagating degrees of freedom, which erroneously can be associated with the pole (\ref{k}). 
 The source of this pole is obviously related to the same topological instanton-like long ranged configurations  (\ref{Q}) saturating the contact term in the topological susceptibility (\ref{divergence}).  
The singular behaviour 
  (\ref{k}) which simply represents  a non-vanishing expectation value (\ref{Q1}), (\ref{E}), obviously implies that the  integral in momentum space around $\mathbf{k}\sim 0$ does not vanish. Indeed, using $k\rightarrow 0$ behaviour (\ref{k}) one arrives to the following relation 
  \be
 \label{k1}
 \frac{1}{e}\oint_{ |\mathbf{k}| \rightarrow 0}  {a^j}(\mathbf{k}) d  {k}_j= \frac{1}{e} \int d^2 \mathbf{k} \left[ \epsilon^{ij} \partial_{ {k}_i}{a_j}(\mathbf{k}) \right]\nonumber\\
 =\theta  \int d^2 \mathbf{k} ~~\partial_{ {k}_i}\left(\frac{{k}_i}{2\pi \mathbf{k}^2} \right) =\theta  \int d^2 \mathbf{k}~\delta^2(\mathbf{k})=\theta ,
 \ee
which  essentially  represents the same well-known statement about  non-vanishing gauge invariant expectation value (\ref{Q1}), (\ref{E}), but written in the different terms involving the auxiliary topological fields in momentum space.

From (\ref{k1}) one can easily recognize that the auxiliary field $\frac{1}{e}{a^j}(\mathbf{k}) $ in momentum space strongly resembles  the Berry's connection (\ref{berry2}), while  $\frac{1}{e}  \epsilon^{ij} \partial_{ {k}_i}{a_j}(\mathbf{k})$ can be thought as the Berry's curvature discussed  previously   in CM physics, see e.g. \cite{Cortijo:2011aa} for review.
\exclude{\footnote{In fact, the behaviour (\ref{k}) corresponds to the following  non-vanishing azimuthal component in Lorentz gauge in momentum space ${a^{\phi}}(\mathbf{k}\rightarrow 0) \rightarrow\frac{e\theta}{2\pi|\mathbf{k}|}$.  This field saturates the integral   
$\frac{1}{e}\oint  {a_{\phi}} d  {k}_{\phi}=\theta$ and   plays the same  role as 
$\tilde{A}_1$ in $A_0=0$ gauge in  notations of ref.\cite{ChenLee}}} 
The fundamental difference between analysis of our system and the computations of the Berry phase  in CM literature  is that 
  the Berry connection   (\ref{berry2}) in CM systems  is  a collective phenomenon with accumulation of   the geometric phase of the band electrons. It is    represented, as a matter of   convenience rather than necessity,   in terms of the emergent gauge field ${\cal{A}}_i(\mathbf{k}) $. In contrast, in our case, 
  the  topological    fields  $ {a_i}(\mathbf{k})$ represent some fundamental (though auxiliary, non-propagating)  fields  describing the ground state of the underlying gauge theory.   
  These fields are  present in the system even without any matter fields.  The topological features of the auxiliary fields in our case  emerge as a result of   the summation  of the topological sectors in path integral formulation  rather than a result  of a complex interaction  of the band electrons in CM systems. 
  
  Nevertheless, as we observed above, there is very strong  mathematical similarity,   between these two, physically very different, entities. These similarities, in particular, include the following features: while  $ {a_i}(\mathbf{k})$ and  ${\cal{A}}_i(\mathbf{k})$ are gauge-dependent objects, the corresponding integrals  (\ref{berry1}) and (\ref{k1}) are gauge invariant (modulo $2\pi$) observables describing the same property related to the   polarization. The $2\pi$ periodicity for all observables in both systems also has very simple physical explanation. For our system the $2\pi$ periodicity follows from the partition function (\ref{Z_2}), (\ref{poisson}), while in  CM context \cite{Cortijo:2011aa,ChenLee} the $2\pi$ periodicity corresponds to the adiabatic   process when the many body 
  wave function  returns to its physically identical (but topologically different) state.  Furthermore, the main features of the systems are formulated in terms of global  rather than local behaviour, as formulae (\ref{berry1}) and (\ref{k1})  suggest. 
One should comment here that  an explicit computations of the Berry's connection for  a specific CM system very often  requires some  tedious microscopical local computations,  though the final result is in fact describes the global behaviour of the system, not sensitive to any local characteristics.

  We conclude this section with  the following  general comment. We have not produced any new physical results in this section as    the relevant questions in  2d QED such as the expectation value of the electric field at non-zero $\theta$ (represented by eqs. (\ref{Q1}), (\ref{E})), or non-dispersive (contact)  contribution to the topological susceptibility (\ref{final}) have been   computed   long time ago\footnote{In particular, formula  (\ref{final})  can be derived  using the Kogut-Susskind ghost formalism \cite{KS}.}. Our contribution in this section is much more modest.  We reproduced    these known results by using a different technique: we expressed  the relevant correlation functions  in  terms of  the  auxiliary topological fields $  {a_i}(\mathbf{k})$. We established the physical meaning of these fields, and argued that these auxiliary objects  play the same role as Berry's connection  ${\cal{A}}_i(\mathbf{k})$ in CM systems. 
  
  As we shall discuss below, the technical tools developed and tested in this subsection  (by reproducing the known results)  will be very  useful  in our study of a  similar phenomena in physically relevant four dimensional Maxwell theory formulated on the torus.  This mathematical similarity occurs  as a result of   dimensional reduction (to be used below)  which essentially translates  the corresponding 4d problems  into 2d analysis developed in present  section.

     \subsection{Auxiliary topological fields  in 4d Maxwell system}\label{4d_auxiliary} 
      We wish to derive the topological action for the 4d Maxwell system by using the same    technique exploited  in previous subsection \ref{2d_auxiliary}. Our starting point is to  insert  the  delta function, similar to eq. (\ref{delta}),  into the path  integral with the field $b^z(\mathbf{x})$ acting as a Lagrange multiplier
 \be
 \label{delta_4d}
 &&\delta \left[B^z(\mathbf{x})-\epsilon^{zjk} \partial_{j} a_k(\mathbf{x})\right]\sim \\
&& \int {\cal{D}}[b_z] e^{ iL_3\beta\int d^2\mathbf{x} ~b_z(\mathbf{x})\cdot  \left[ B^z(\mathbf{x})-\epsilon^{zjk} \partial_{j} a_k(\mathbf{x})\right]} \nonumber
  \ee
 where $B^z(\mathbf{x})$   in this formula is treated as the original expression for the field operator entering the action (\ref{action4d}),  
		including all  classical k-instanton configurations (\ref{topB4d},\ref{action4d2}) and quantum fluctuations surrounding these classical configurations. In other words, we treat  $B^z(\mathbf{x})$ as fast degrees of freedom. At the same time  $a_k(\mathbf{x})$  is treated as a slow-varying   external source effectively describing the large distance physics  for a given instanton configuration.  Our task now is to integrate out the original   fast ``fluxes" (\ref{topB4d},\ref{action4d2})  and describe the large distance physics in terms of slow varying fields $b_z(\mathbf{x}), a_k(\mathbf{x})$ in form of the effective action similar to (\ref{S_tot}) derived for 2d system. The physical meaning of these formal manipulations is explained in the previous section  \ref{2d_auxiliary} after eq. (\ref{delta}), and we shall not repeat it here. 
		
		To proceed with computations, we use the same procedure by summation over k-instantons 
as   described in section \ref{4d}.  The only new element 
in comparison with the previous computations is that  the fast degrees of freedom must be integrated out in the presence of the new slow varying  background fields   $b_z(\mathbf{x}), a_k(\mathbf{x})$ which appear in  eq. (\ref{delta_4d}). Fortunately, the computations can be easily performed if one notices that the background field $b_z(\mathbf{x})$ enters eq. (\ref{delta_4d}) exactly in the same manner as external magnetic field   enters  (\ref{Z_eff}).
Therefore, assuming that $b_z(\mathbf{x}), a_k(\mathbf{x})$ are slow varying background fields we arrive to the following 
  expression for the   partition function for our   $\cal{TV}$ system:
  \be
\label{Z_4d}
{\cal{Z}}_{\rm top} (\tau, \theta_{\rm eff}) =\sqrt{\pi\tau}\sum_{k\in {\mathbb{Z}}} \int {\cal{D}}[b_z]  {\cal{D}}[a]  e^{-S -S_{\rm top}  }
\ee
where quadratic action $ S[b_z, a_k]$ is defined as 
\be
\label{Z_BF_4d}
 S[b_z, a_k]= \pi^2\tau\int_{{\mathbb{T}}_2} \frac{d^2 \mathbf{x}}{L_1L_2}
\left(k+\frac{\phi (\mathbf{x})+\theta_{\rm eff}}{2\pi}\right)^2, 
\ee
while the topological term  $S_{\rm top}[b_z, a_k] $ in eq. (\ref{Z_4d}) reads  
\be
\label{S_top_4d}
 S_{\rm top}[b_z, a_k] 
 =  i L_3\beta\int_{{\mathbb{T}}_2} d^2\mathbf{x} \left[b_z(\mathbf{x}) \epsilon^{zjk} \partial_{j} a_k(\mathbf{x})\right]. 
\ee
In formula  (\ref{Z_BF_4d})    we rescale  the slow varying background auxiliary $b_z$ field such that   $\phi (\mathbf{x})\equiv eL_1L_2 b_z (\mathbf{x})$. Parameter   
    $\theta_{\rm eff}\equiv eL_1L_2 B_z^{\rm ext}$
represents the external magnetic field 
  while  ${\mathbb{T}}_2$ represents the  two torus defined on $(1,2)$ plane.   
  
  The topological action (\ref{S_top_4d}) in all respects is very similar to the topological action derived for 2d system (\ref{S_top}). Therefore, we anticipate that all consequences discussed in previous section \ref{2d_auxiliary} for 2d system will have their analogous in 4d system, including the relation between the Berry's connection and auxiliary fields in momentum space in the IR at $\mathbf{k}\rightarrow 0$.

Before we proceed with computations  to establish such a connection, we 
want to make the following comment. The topological term (\ref{S_top_4d}) which emerges as an effective description of our system 
is in fact a   Chern-Simons like  topological action. 
In our simplified setting   we limited ourself by considering the fluxes along $z$ direction only. It is naturally to assume that a more general construction would include fluxes   in all three directions  which would  lead to a generalization of  action   (\ref{S_top_4d}). Therefore, it is quite natural to expect  that   the action in this case would   assume a  Chern-Simons like form $i \beta\int_{{\mathbb{T}}_3} d^3\mathbf{x} \left[\epsilon^{ijk} b_i(\mathbf{x}) \partial_{j} a_k(\mathbf{x})\right]$ which replaces (\ref{S_top_4d}).
A similar structure in CM systems is known to describe a  topologically ordered phase.  	Therefore, it is not really a surprise that we observed  in \cite{Zhitnitsky:2013hba} some signatures of the topological order  in the Maxwell system defined  on a compact manifold. The emergence of the topological Chern-Simons action (\ref{S_top_4d})  further supports this basic claim that the Maxwell system  on a compact manifold belongs to  a topologically ordered phase as the auxiliary topological fields entering (\ref{S_top_4d}) play the same role as the Berry's connection in topologically ordered CM systems.

Now we can follow the same procedure which we tested  for 2d system in section \ref{2d_auxiliary}
to compute the expectation value  of the magnetic field at non-vanishing $\theta_{\rm eff}$.
The corresponding result is known \cite{Cao:2013na}: it has been derived by  using conventional computation of the path integral by summation  over all ``instanton-fluxes". Our goal now is to reproduce this result by using the auxiliary topological fields governed by the action (\ref{S_top_4d}). We follow the same procedure as before  and define the induced magnetic field in the system in the conventional way
\be
 \label{B1}
 && \left<  B^z_{\rm ind}( \tau, \theta_{\rm eff}) \right> =-\frac{1}{\beta V}\frac{\partial \ln{{\cal{Z}}_{\rm top}} ( \tau, \theta_{\rm eff})}{\partial B_{\rm ext}}\\
 && =\frac{2\pi}{eL_1L_2}\left< k+\frac{\theta_{\rm eff}+\phi (\mathbf{x})}{2\pi}\right> ,\nonumber
  \ee
  where the last expectation value must be evaluated using the partition function (\ref{Z_4d}). 
  The corresponding Gaussian integral over auxiliary $\phi (\mathbf{x})$ field can be easily executed with the result
  \be
 \label{B2}
 \left<  B^z_{\rm ind}( \tau, \theta_{\rm eff}) \right> =  \lim_{\mathbf{k}\rightarrow 0} \int \frac{d^2\mathbf{x}}{L_1L_2}e^{i\mathbf{kx}}  \left< -i \epsilon^{zjk} \partial_{j} a_k(\mathbf{x})\right>  ,~~~~~
  \ee
        where the corresponding expectation value $\left<... \right>$ should be computed using  the following 
        partition function determined by the    action $S_{\rm tot}[a_k]$ (which  includes both: the quadratic and topological terms), 
  \be
\label{Z_tot_4d}
&& S_{\rm tot}[a_k]=  \frac{L_3\beta}{2}\int_{{\mathbb{T}}_2}  d^2 \mathbf{x}  \left(\epsilon^{zjk} \partial_{j} a_k(\mathbf{x})\right)^2 \\&&-i  \left(2\pi k+\theta_{\rm eff}\right)\frac{L_3\beta}{eL_1L_2}\int_{{\mathbb{T}}_2}  d^2 \mathbf{x} 
  \left(\epsilon^{zjk} \partial_{j} a_k(\mathbf{x})\right). \nonumber
 \ee
  The  path integral  integral (\ref{B2}) is gaussian, and can be executed by a conventional shift of variables in the action $S_{\rm tot}[a_k]$ defined  by (\ref{Z_tot_4d})
  \be
   \left(\epsilon^{zjk} \partial_{j} a'_k(\mathbf{x})\right)= \left(\epsilon^{zjk} \partial_{j} a_k(\mathbf{x})\right)-i\frac{\left(2\pi k+\theta_{\rm eff}\right)}{eL_1L_2}.
  \ee
Exact evaluation of the gaussian  path integral  (\ref{B2}) with action (\ref{Z_tot_4d})  leads to the following final result for  $ \left<  B^z_{\rm ind} \right>$  
\be
\label{B3}
 \left<  B^z_{\rm ind} \right>=\frac{2\pi}{eL_1L_2}\frac{\sqrt{\pi\tau}}{{\cal{Z}}_{\rm top}}
 \sum_{k \in \mathbb{Z}} \left(k+\frac{\theta_{\rm eff}}{2\pi}\right) e^{-\pi^2\tau \left(k+\frac{\theta_{\rm eff}}{2\pi}\right)^2},~~~~~~
 \ee
 where the  partition function  ${\cal{Z}}_{\rm top}$ for our   $\cal{TV}$ system in this formula\footnote{We note that $k$-independent numerical factor
 $\sqrt{\pi\tau}$  enters both  equations: (\ref{Z_eff}) and (\ref{B3}). This numerical factor simply represents our  normalization's convention, and does not affect  the computations of any expectation values, such as (\ref{B3}). Our normalization corresponds to    the following behaviour of  the topological   partition function:  ${\cal{Z}}_{\rm top}\rightarrow 1$ in the limit $L_1L_2\rightarrow\infty$. Such a convention    corresponds to the geometry of the  original  Casimir setup experiment, see \cite{Cao:2013na}  for the details.} is   determined  by eq. (\ref{Z_eff}). 
 As expected, the expression (\ref{B3})  exactly reproduces the corresponding formula derived in ref\cite{Cao:2013na}    by explicit summation over fluxes-instantons. We reproduced the results of  ref\cite{Cao:2013na}  using drastically different technique as our   computations (\ref{B3}) in this section  are based on calculation    of the  path integral defined by the partition function (\ref{Z_4d}) formulated in terms of the auxiliary topological fields $b_z(\mathbf{x}), a_i(\mathbf{x}) $. Agreement between the two approaches obviously supports the  consistency of our formal manipulations with the path integral and auxiliary fields. 
 
 Important  new point (which could not be seen within  computational technique of  ref\cite{Cao:2013na}) is the expression  (\ref{B2}) for the induced field $\left<  B^z_{\rm ind} \right>$ in terms of the auxiliary object $a_i(\mathbf{x})$. As we shall see in a moment  precisely this connection allows us to identify  the auxiliary topological non-propagating field $a_i(\mathbf{k})$ in momentum space with the Berry's connection  ${\cal{A}}_i (\mathbf{k})$ from section \ref{2d-berry} as both entities have very similar properties. 
 
 Before we proceed to establish  such a connection, we would like to make few  comments.
 First, as one can see from (\ref{B3}) the  expression for  $\left<  B^z_{\rm ind} \right>$ accounts for the total field in the system,  including the external field as well as the induced  field due to the interference of the external field with the topological fluxes  (\ref{topB4d}). 
 However,  in the absence of the external field $\theta_{\rm eff}\equiv eL_1L_2 B_{\rm ext}^z=0$ the contributions to the expectation value (\ref{B3})  from the fluxes  with positive and negative signs cancel each other,  and $ \left<  B^z_{\rm ind} \right>$ vanishes. 
 For $\theta_{\rm eff}\neq 0$ the cancellation does not hold, and the   field $ \left<  B^z_{\rm ind} \right>\neq 0$ will be obviously induced. 
 
The effect must vanish when the tunnelling transitions due to the fluxes  are suppressed at $e\rightarrow 0$ which corresponds to large $\tau\gg 1$.   It is very instructive to see  how it happens. The corresponding expression which is valid for $\tau\gg 1$ and small but finite $\theta_{\rm eff}\ll 1$
reads
\be
\label{B4}
 \left<  B^z_{\rm ind} \right>=\frac{\theta_{\rm eff}}{eL_1L_2}\left[1-\frac{4\pi e^{-\pi^2\tau}\sinh (\pi\tau \theta_{\rm eff}) }{\theta_{\rm eff}} \right].
\ee
  One can explicitly see from eq. (\ref{B4}) that the tunnelling effects are suppressed in large $\tau$ limit, and magnetic field in this case in the system is entirely determined by external source $ \left<  B^z_{\rm ind} \right> \rightarrow B_{\rm ext}^z$ at $\tau\gg 1$, as expected. 
  
 The  key point for the present analysis is the expression (\ref{B2}) for $ \left<  B^z_{\rm ind} \right>$ in terms of the auxiliary fields $a_k(\mathbf{x})$. This formula in all respects is very similar to expression (\ref{Q1}) previously analyzed in 2d system. One can follow the same logic of that analysis to arrive to the conclusion that the auxiliary field $a_k(\mathbf{x})$ can be thought as the Berry's connection (similar to (\ref{k})  from 2d analysis) with the following singular behaviour at small  $\mathbf{k}\rightarrow 0$  in momentum space, 
  \be
 \label{B5}
&& a^i(\mathbf{k}\rightarrow 0)\equiv \frac{1}{L_1L_2}\int \frac{d^2\mathbf{x}}{2\pi} e^{i\mathbf{kx}} a^i(\mathbf{x}) \\
&&=\left<  B^z_{\rm ind} \right>\frac{\epsilon^{zij}   {k}_j}{2\pi \mathbf{k}^2} \Rightarrow \left( \frac{\theta_{\rm eff}}{eL_1L_2}\right)\frac{\epsilon^{zij}   {k}_j}{ 2\pi\mathbf{k}^2}, \nonumber
 \ee
where in the last line we use the asymptotical behaviour (\ref{B4}) which is valid for large $\tau\gg 1$. 

The behaviour (\ref{B5}) also suggests   that $\epsilon^{zij} \partial_{k_i}a_j(\mathbf{k})\sim \delta^2(\mathbf{k})$ plays the same role as the Berry's curvature in CM physics, see (\ref{berry2}) and \cite{Cortijo:2011aa} for review. One should emphasize that  these similarities in  the  IR behaviour  in two very different systems  should not be considered as a pure  mathematical curiosity. In fact, there is a  very deep physical reason why  these two, naively unrelated, entities, must behave very similarly in the IR. Indeed, 
 as it is known  the Berry's phase in CM systems  effectively describes the variation of the $\theta$ parameter $\theta\rightarrow \theta-2\pi P$ as a result of coherent influence of strongly interacting fermions which polarize the system, i.e. $P=\pm 1/2$, see e.g. \cite{ChenLee}. 
The auxiliary topological field $a^i(\mathbf{x})$  in our  $\cal{TV}$ system with similar IR behaviour essentially describes  the same physics.  To be more precise, the interference between the external magnetic field  and fluxes  lead to the magnetic polarization formulated in terms of $a^i(\mathbf{x})$ fields,    similar to the  generation of polarization $P=\pm 1/2$ in CM systems   expressed in terms of  the Berry's connection ${\cal{A}}^i(\mathbf{x})$.    Our key observation is that  the polarization features  of the  $\cal{TV}$ system in our    case  are   represented by eqs. (\ref{B2}), (\ref{B5}). These equations     play the same role as equations (\ref{berry1}),  (\ref{berry2})  in CM systems.
 
 This close analogy (mathematical and  physical), in fact, may have some profound observational and experimental  consequences
 as an electrically charged probe inserted into our system characterized by   $\theta_{\rm eff}$ would behave very much in the same way as a probe inserted into a CM system characterized by non-vanishing Berry's phase. In other words, our  $\cal{TV}$ system must demonstrate a number of unusual features which are   typical for topologically ordered phases  in CM systems. One of such properties, the degeneracy of the system, which can not be  described in terms of  any local operator (but rather is characterized by 
 a non-local operator) has been already established \cite{Zhitnitsky:2013hba}. It must be other  interesting experimentally observable effects  in 4d Maxwell theory, similar to  a number of profound effects  which are known to occur in topologically ordered phases  in CM systems \cite{Wen:1989iv,Wen:1990zza,Moore:1991ks,BF,Cho:2010rk,Wen:2012hm,Sachdev:2012dq, Cortijo:2011aa, Volovik:2011kg}. We leave this subject for future studies.

      \section{Conclusion and Future Directions }\label{conclusion}
         In this work we discussed a number of very unusual features   exhibited by    the   Maxwell theory formulated on  a 4 torus, which was coined the  topological vacuum ($\cal{TV}$).  All these  features are originated from the topological portion of the partition function ${\cal{Z}}_{\rm top}(\tau, \theta_{\rm eff})$
  and   can not be formulated in terms of  conventional  $E\&M$  propagating photons  with two physical transverse polarizations. In different words, all effects discussed in this paper have a non-dispersive nature. 
  
             The computations of the present work along with previous calculations of refs.\cite{Cao:2013na,Zhitnitsky:2013hba} imply  that the extra energy (and entropy),   not associated   with any physical propagating degrees of freedom,  may emerge  in the  gauge  systems if some conditions are met. This fundamentally new type  of  energy    emerges as a result of dynamics of pure gauge configurations at arbitrary  large distances. This unique feature of the system when an extra energy is not related to any physical propagating degrees of freedom was the main  motivation for a proposal   \cite{Zhitnitsky:2011tr,Zhitnitsky:2011aa,Zhitnitsky:2013pna}  that the  vacuum energy of the Universe may have, in fact,  precisely such non-dispersive  nature\footnote{There are two instances in evolution of the universe when the vacuum energy plays a crucial  role.
             First instance   is identified with  the inflationary epoch  when the Hubble constant $H$ was almost constant which corresponds to the de Sitter type behaviour $a(t)\sim \exp(Ht)$ with exponential growth of the size $a(t)$ of the Universe. The  second instance when the vacuum energy plays a dominating role  corresponds to the present epoch when the vacuum energy is identified with the so-called dark energy $\rho_{DE}$
   which constitutes almost $70\%$ of the critical density. In the proposal \cite{Zhitnitsky:2011tr,Zhitnitsky:2011aa,Zhitnitsky:2013pna}  the vacuum energy density can be estimated as $\rho_{DE}\sim H\Lambda^3_{QCD}\sim (10^{-4}{\rm  eV})^4$, which is amazingly  close to the observed value. }. 
This proposal when an extra energy   can not be associated with any propagating particles  should be contrasted with a conventional description when an extra vacuum energy is always associated with some   new physical propagating degree of freedom, such as inflaton. 
   
   The main motivation for  the present studies     is to test these ideas (about  fundamentally new type of vacuum energy)  using a simple  quantum field theory (QFT) setting  which nevertheless preserves  the crucial  element, the degeneracy of the topological sectors,  responsible for this novel type of energy. This simplest possible setting 
   can be  realized in  the   Maxwell theory formulated on  a 4 torus.  Most importantly, the effect with this simplest   setting can be, in principle,
   tested in a tabletop experiment if the corresponding boundary conditions can be somehow imposed in a real physical systems. Otherwise, our construction should be considered as the simplest possible 4d QFT model when an  extra vacuum energy is generated. The crucial point is that this extra vacuum energy can not be associated with any physical propagating degrees of freedom, as argued in the present work.    
   
    Essentially, the  proposal  \cite{Zhitnitsky:2011tr,Zhitnitsky:2011aa,Zhitnitsky:2013pna}  identifies the observed vacuum  energy with the Casimir type energy, which however is originated not from dynamics of the physical massless propagating degrees of freedom, but rather, from the dynamics of the topological sectors  which  are always present in gauge systems, and which are highly sensitive to arbitrary large distances. 
The vacuum energy   in this case can be formulated in terms of the auxiliary topological fields which are
similar in spirit to   $b_z (\mathbf{x}), a_k (\mathbf{x})$ fields from (\ref{S_tot}), (\ref{S_top_4d})   and which effectively describe the dynamics of the topological sectors in the expanding background \cite{Zhitnitsky:2013pna}.  As we discussed in length in section \ref{4d_auxiliary} these auxiliary topological fields play the same role as the Berry's connection in CM systems.  The $b_z (\mathbf{x}), a_k (\mathbf{x})$ fields do not propagate, but they do contribute to the vacuum energy.  It would be very exciting if this new type of the vacuum energy not associated   with    propagating particles could be experimentally measured in a laboratory, as we advocate in this work.
        
      Aside from testing the ideas on vacuum energy of the Universe, the Maxwell system studied in the present work is interesting system on its own. Indeed, being a ``free" Maxwell theory, it nevertheless  shows a number of very unusual features which are normally attributed to a CM system in a topologically ordered phase. In particular, it shows the degeneracy of the system which 
      can not be detected by any local operators, but characterized by a non-local operator \cite{Zhitnitsky:2013hba}. Furthermore, in the present work we argued that the auxiliary topological fields  $b_z (\mathbf{x}), a_k (\mathbf{x})$ fields  in 4d Maxwell system behave very much in the same way as the Berry's connection in CM systems. More than that,  a charged probe particle inserted into our system would feel the topological features of the $b_z (\mathbf{x}), a_k (\mathbf{x})$ fields  in the same way as  a probe inserted into a CM system characterized by  a nontrivial  Berry's connection ${\cal{A}}^i(\mathbf{x})$.
     
       Therefore, it would be very exciting if one could find a   system where a charged probe inserted into our 4-torus  (filled by vacuum) would behave similarly to a probe inserted into a  much more complicated CM system, where the corresponding nontrivial Berry's connection is emergent as a  result of a coherent many body physics. 
       
       A simplest possible setup we can imagine is as  follows. Normally, in condensed matter literature one  considers a junction between a conventional  insulator ($\cal{I}$) and topological insulator ($\cal{TI}$).  One can also consider the $\cal{TI}$  which is sandwiched between two conventional insulators, i.e. one can consider a  system like  $\cal{I-TI-I}$.  
       Our claim  essentially is that the  $\cal{TI}$  in this system can be replaced by the   $\cal{TV}$ configuration considered  in this work.
      In other words, one considers a system like $\cal{I-TV-I}$. Our claim is that this system would behave very much as the   $\cal{I-TI-I}$,
      because  $\cal{TV}$ behaves very much in the same way as a $\cal{TI}$  as advocated in this work\footnote{The conducting (or superconducting)  edges of the  $\cal{TV}$ portion in $\cal{I-TV-I}$ sandwich in form of a closed circle support  the periodic boundary conditions up to the large gauge transformations.  This freedom in terms of large gauge transformation, or what is the same, a finite probability to form the ``instanton-fluxes" in the bulk of the system,  eventually leads to the emergence of the partition function (\ref{Z4d}) with all its consequences discussed in the present work. Instanton fluxes  (\ref{topB4d})   also satisfy the boundary condition $B_{\perp}=0$ on the superconducting edges.}. These similarities  include such nontrivial features  as  the degeneracy (characterized by a non-local operator), the Berry's connection and the presence of the effective $\theta_{\rm eff}$ state, among   many others things. Therefore, it is naturally to expect that  while     $\cal{ I-TI-I}$ and $\cal{I-TV-I}$  systems are very different in composition, the behaviour of these systems will be very much the same in the IR.  We leave this exciting  subject on possible applications of our  $\cal{TV}$ system, which we believe belongs to a topologically ordered phase,  for future investigations.

 \section*{Acknowledgements} 
  I am thankful to Maria Vozmediano for useful and stimulating discussions on possibilities to test the ideas discussed in the present work in a real tabletop experiment.  I am also thankful to   other participants of 
 the program  ``Quantum Anomalies, Topology and    Hydrodynamics", 
 Simons Center for Geometry and Physics, Stony Brook, February -June, 2014, where this work was presented. 
 Finally, I am thankful to Alexei  Kitaev for long and interesting discussions related to the subject of the present work.
  This research was supported in part by the Natural Sciences and Engineering Research Council of Canada.


\end{document}